\documentclass[12pt,a4paper]{revtex4}
\usepackage[utf8x]{inputenc}
\usepackage{ucs}
\usepackage[english]{babel}
\usepackage{amsmath}
\usepackage{amsfonts}
\usepackage{amssymb}
\usepackage{graphicx}
\usepackage[left=2cm,right=2cm,top=2cm,bottom=2cm]{geometry}

\begin{document}
\title{Analytical solution of coupled self--consistency and linearised Usadel equations for the dirty superconductors at $T_c$ and with the proximity effect}

\author{S.S. Seidov}
\affiliation{HSE University, Moscow, Russia}
\author{N.G. Pugach}
\affiliation{HSE University, Moscow, Russia}

\begin{abstract}
In this manuscript we consider a superconducting film in the vicinity of the critical temperature and presence of the proximity effect. We analytically solve the corresponding linearised Usadel equation and the self--consistency equation, defining the critical temperature. This is a system of coupled differential and integral equations for the anomalous Green function and the order parameter of the superconductor. The proximity effect defines the boundary conditions. The formal solution of the system is found for the general case of the linearised boundary conditions defined by the proximity effect, reducing the set of equations to an eigenvalue problem. The latter defines the critical temperature of the superconducting phase transition and the spatial distributions of the anomalous Green function and the superconducting order parameter.
\end{abstract}
\maketitle

\section{Introduction}
Finite--sized dirty superconductors present significant interest from both fundamental and practical points of view. When the superconductor is brought in contact with different materials, the properties of electronic subsystems both in the superconductor and the second material are changed due to mutual influence --- this is the so--called proximity effect \cite{de_gennes_boundary_1964, buzdin_proximity_2005, eschrig_spin-polarized_2015}. This allows to create uncommon electronic states in both materials, such as suppression and induction of superconductivity, and also triplet electron pairing \cite{bergeret_odd_2005, eschrig_general_2015} and topological states \cite{Leijnse_2012, hassler_majorana_2014, seleznev_density_2025} at the proximity with magnetic materials. These effects are applicable in the field of superconducting spintronics \cite{linder_superconducting_2015, leksin_peculiarities_2013, leksin_boosting_2016, kamashev_superconducting_2019, gordeeva_record_2020, heim_ferromagnetic_2013}. Given that the proximity effect shifts the critical temperature of the phase transition, it can be used for designing superconducting spin valves. In particular, by controlling e.g. the magnetization of the second material, the superconductor can be switched between the superconducting and normal states, thus opening and closing the spin valve. This motivates the current study, as one is interested in the properties of the critical temperature of the superconducting layer, when designing the superconducting spin valve, as well as the spatial distributions of the order parameter and the anomalous Green function in the superconductor.

The dirty superconductor is described by two equations: the Usadel equation \cite{usadel_generalized_1970} and the self--consistency equation \cite{tinkham_introduction_2015}. The first one defines the spatial distribution of the normal and anomalous Green functions in the superconductor. The second one defines the spatial distribution of the order parameter and the critical temperature of the phase transition. Together they form a system of a coupled nonlinear differential equation and an integral equation. The last one contains the integral over the energy parameter while the second order differential Usadel equation has the coordinate derivatives. Its numerical solution requires computationally expensive iterative procedure, e.g. described in works \cite{tumanov_2024}. However, if the temperature is close to the critical temperature, the Usadel equation can be linearised and becomes a second order linear equation. It still remains coupled to the integral equation for the order parameter, but extensive analytical study can be performed, which is the goal of our study. 

Although in principle the Ginzburg--Landau equation is also applicable for the order parameter calculation, the choice of related boundary conditions remains problematic, especially in cases of proximity with ferromagnets both metallic and insulating, spin-orbit interaction at the interface, and other complicated and/or spin--active proximized materials \cite{buzdin_proximity_2005}. The transformation of the microscopic boundary conditions into the boundary condition for Ginzburg--Landau equation was developed in \cite{ivanov_boundary_1981}, but this approach is valid only in the limit $\xi \ll d \ll \xi(T)= (\pi\xi/2) \sqrt{1-T/T_c}$, where $\xi$ is the microscopic coherence length, $d$ is the thickness of the superconducting film and $\xi(t)$ is the Ginzburg--Landau coherence length.

In the present manuscript we express the solution of the Usadel equation in the integral form, substitute it in the self--consistency equation and thus decouple two equations. Similar approach was proposed in \cite{fominov_nonmonotonic_2002, antropov_experimental_2013}. In contrast to \cite{fominov_nonmonotonic_2002}, we represent the integral kernel in the self--consistency equation as an infinite sum. The leading term in the sum corresponds to the homogeneous solution, so one can consider perturbative corrections due to proximity effect. Additionally, the integral equation is reduced to an eigenvalue problem for a certain matrix, matrix elements of which are obtained. This reduction to a matrix eigenvalue problem was also made in \cite{antropov_experimental_2013} for the particular case of a ferromagnet--superconductor--ferromagnet trilayer. Our study considers an arbitrary junction with Robin boundary conditions and a more extensive study of the arising mathematical structure. From physical point of view, the results in the present manuscript are equivalent to results in \cite{fominov_nonmonotonic_2002, antropov_experimental_2013}, when applied to corresponding physical systems. This is due to mathematical equivalence of the considered physical problems. The critical temperature of the superconductor is then defined by condition of existence of the nonzero solution of said eigenvalue problem. In general case, the eigenvalue problem has to be solved numerically, however this is significantly less computationally expensive for $T_c$ calculation, than performing a usual iterative procedure. Finally, the solution defines the spatial distributions of the anomalous Green function and the order parameter in the superconductor. The equation for the critical temperature of the proximity suppressed superconducting film is also obtained. We illustrate the method by obtaining the self--consistency equation for the superconductor with proximity effect at a single border. We derive the corresponding self--consistency equation in the case of the thin superconductor and numerically find corrections to the bulk critical temperature. The dependences of the critical temperature on the strength of the proximity effect for different thickness of the superconducting film are presented.

\section{Statement of the problem}
We consider a finite--sized superconductor of length $d$ in the dirty limit, brought in contact with some other materials such as another superconductor, normal non-superconducting metal, magnet, etc at $x = 0$ and $x = d$. The goal is to find the critical temperature $T_c$ of the superconducting phase transition, the spatial distributions of the anomalous Green function and the order parameter. The order parameter and the critical temperature are connected via the self--consistency equation (we set $\hbar = k_B = 1$ through out the paper) 
\begin{equation}\label{eq:SC}
\Delta(x) = \frac{U}{2} \int\limits_0^{\omega_D} \Re f_s(x) \tanh \left(\frac{\varepsilon}{2T_c} \right) d \varepsilon.
\end{equation}
Here $f_s(x)$ is the anomalous Green function and for the dirty superconductor at $T = T_c$ it obeys the linearised Usadel equation \cite{buzdin_proximity_2005}
\begin{equation}\label{eq:Usadel}
\begin{aligned}
&f_s''(x) -k^2 f_s(x) = \frac{2 \Delta(x)}{i D}\\
&k^2 = \frac{2 i \varepsilon}{D}.
\end{aligned}
\end{equation}
The proximity effect, induced via the contact with another materials at the border, defines the boundary conditions \cite{zaitsev_quasiclassical_1984, kurpianov_influence_1988, linder_quasiclassical_2022, eschrig_general_2015}.  Additionally, the boundary conditions depend on internal properties of the superconductor, such as, for example, presence of spin--orbit coupling \cite{bergeret_spin-orbit_2014, kokkeler_universal_2025}. We consider the linearised boundary conditions in form
\begin{equation}\label{eq:BC}
\begin{aligned}
&f_s'(0) = \alpha_0 (\varepsilon) f_s(0) &f_s'(d) = \alpha_d (\varepsilon) f_s(d).
\end{aligned}
\end{equation}
The coefficients $\alpha_{0,l}$ are defined by the exact type of the materials in contact with the superconductor. This simple form is rather general while we consider a usual BCS s--wave superconductor, where the order parameter is created only by singlet superconducting correlations. The singlet to triplet conversion at the ferromagnetic interface, possibly with a spin--orbit interaction, together with the drainage of the Cooper pairs into the neighbour metal due to the proximity effect can be considered by the right choice of the parameter alpha for the only singlet component of the anomalous Green function \cite{fominov_nonmonotonic_2002}. If the materials in proximity with the superconductor are conductive and the anomalous Green function is not zero in the corresponding regions, one can solve first the linearized Usadel equation in this regions and reduce the boundary conditions for the Green function in the superconductor to form (\ref{eq:BC}), an example of this calculation for the contact with a semiinfinite normal metal is conducted in Appendix \ref{sec:appBC}. So, from a mathematical point of view, we must solve a system of coupled integral and differential equations. 

\section{Formal solution}
\subsection{Solution of the Usadel equations}
First, we have to express the solution of the Usadel equation (\ref{eq:Usadel}) as an integral of $\Delta(x)$ with some integral kernel. This is done by applying a usual Green function technique of solving nonhomogeneous differential equations \cite{duffy_greens_2001}. In particular,
\begin{equation}\label{eq:f_sol}
f_s(x) = a_+(\varepsilon) e^{k x} + a_-(\varepsilon) e^{-k x} + \frac{2}{i D}\int\limits_0^d H(x, x'; \varepsilon) \Delta(x') dx'.
\end{equation}
Here $H(x, x'; \varepsilon)$ is the Green function of the differential operator $\hat L = \partial_x^2 - k^2(\varepsilon)$ with homogeneous boundary conditions. It can be thus presented as a sum in terms of eigenfunctions of $\hat L$:
\begin{equation}\label{eq:H0_eigen}
H(x, x'; \varepsilon) = \sum_{n = 0}^\infty \frac{\psi_n(x) \psi_n(x')}{h_n}.
\end{equation}
Here $\psi_n(x)$ and $h_n$ are solutions of the equation
\begin{equation}
\begin{aligned}
&\psi''_n(x) - k^2 \psi_n(x) = h_n \psi_n(x)\\
&\psi'_n(0) = \psi'_n(d) = 0.
\end{aligned}
\end{equation}
From this we find 
\begin{equation}\label{eq:psi}
\begin{aligned}
&\psi_n(x) = \sqrt\frac{2}{d} \cos \left(\frac{\pi n x}{d} \right)\\
&h_n= - k^2 - \frac{\pi^2 n^2}{d^2} = - \frac{2 i \varepsilon}{D}  - \frac{\pi^2 n^2}{d^2}.
\end{aligned}
\end{equation}

Next, from the boundary conditions (\ref{eq:BC}) one finds the coefficients $a_\pm (\varepsilon)$, which turn out to be in form
\begin{equation}\label{eq:apm}
a_\pm (\varepsilon) = \sum_n \lambda_n^\pm(\varepsilon) \int\limits_0^d \psi_n(x')  \Delta(x') dx.
\end{equation}
For derivation and explicit expressions for $\lambda_n^\pm(\varepsilon)$ see Appendix \ref{sec:appLambda}. Finally, the solution of the Usadel equation is brought in form
\begin{equation}\label{eq:fs_H}
\begin{aligned}
f_s(x) &= \int\limits_0^d \sum_n \left\{\lambda_n^+(\varepsilon) e^{k x} + \lambda_n^-(\varepsilon) e^{-k x} + \frac{2\psi_n(x)}{i D h_n} \right\} \psi_n(x')\Delta(x') d x' = \\
&= \sum_n \varphi_n(x, \varepsilon) \int\limits_0^d  \psi_n(x')\Delta(x') d x'.
\end{aligned}
\end{equation}
Here we have denoted the function in curly brackets as $\varphi_n(x, \varepsilon)$.

\subsection{Solution of the self--consistency equation}
So, in equation (\ref{eq:fs_H}) we have expressed the anomalous Green function as an integral of the order parameter $\Delta(x)$ with a separable kernel. This expression, when substituted in the self--consistency equation (\ref{eq:SC}), gives an integral equation for the order parameter:
\begin{equation}\label{eq:Delta_int_eq}
\Delta(x) = \frac{U}{2} \sum_n \int\limits_0^{\omega_D} \tanh \left(\frac{\varepsilon}{2T_c} \right)  \Re \varphi_n(x, \varepsilon) d \varepsilon \int\limits_0^d  \psi_n(x')\Delta(x') d x'.
\end{equation}
This a homogeneous Fredholm integral equation of the second kind and in essence is an eigenvalue problem for the linear operator with a corresponding integral kernel \cite{polyanin_handbook_1998}. Conveniently, the functions $\psi_n(x)$ do not depend on $\varepsilon$, so they can be taken out of the corresponding integral and integrating with respect to $\varepsilon$ this equation is brought in form
\begin{equation}\label{eq:Delta_expansion}
\begin{aligned}
&\Delta(x) = \frac{U}{2} \sum_n \Phi_n(x; T_c) \int\limits_0^d  \psi_n(x')\Delta(x') dx'\\
&\Phi_n(x; T_c) = \int\limits_0^{\omega_D} \Re\varphi_n(x, \varepsilon) \tanh \left(\frac{\varepsilon}{2T_c} \right) d \varepsilon.
\end{aligned}
\end{equation}
Multiplying both sides of the equation by $\psi_m(x)$ and integrating over $x$ we find
\begin{equation}\label{eq:psi_Delta}
\int\limits_0^l \psi_m(x) \Delta(x) dx = \frac{U}{2} \sum_n \int\limits_0^d \psi_m(x) \Phi_n(x; T_c) dx\int\limits_0^d  \psi_n(x')\Delta(x') dx'.
\end{equation}
This is an eigenvalue problem for a matrix, which can be seen if one denotes 
\begin{equation}\label{eq:cK}
\begin{aligned}
&c_n = \int\limits_0^d \psi_n(x) \Delta(x) dx \\
&K_{mn}(T_c) = \int\limits_0^d \psi_m(x) \Phi_n(x; T_c) dx.
\end{aligned}
\end{equation}
The matrix elements $K_{mn}$ are known because the functions $\psi_m(x)$ and $\Phi_n(x; T_c)$ are defined by equations (\ref{eq:psi}) and (\ref{eq:Delta_expansion}) accordingly. The unknown coefficients $c_n$ are defined by equation (\ref{eq:psi_Delta}) which becomes
\begin{equation}
c_m = \frac{U}{2} \sum_n K_{mn}(T_c) c_n.
\end{equation}
This is an eigenvalue problem for the matrix $\hat K(T_c)$ with matrix elements $K_{mn}(T_c)$. The characteristic equation
\begin{equation}\label{eq:det_Tc}
\left|\hat K(T_c) - \frac{2}{U} \right| = 0
\end{equation}
determines the critical temperature $T_c$ and the components $c_n$ of the corresponding eigenvector determine expansion of the order parameter $\Delta(x)$ in terms of functions $\Phi_n(x)$ according to equation (\ref{eq:Delta_expansion}). 

To summarize, we propose the following procedure for numerical calculation of the critical temperature:
\begin{enumerate}
\item Define functions $\varphi_n(x, \varepsilon)$ given by equation (\ref{eq:fs_H}) and the coefficients $\lambda^\pm_n(\varepsilon)$ given by (\ref{eq:lambda})
\item Define the functions $\Phi_n(x; T_c)$ given by (\ref{eq:Delta_expansion})
\item Define the matrix $\hat K(T_c)$ according to equation (\ref{eq:cK})
\item Find such value of $T_c$ at which the equation (\ref{eq:det_Tc}) is satisfied
\end{enumerate}
In principle, the matrix $\hat K(T_c)$ is of infinite size, so one should impose a cutoff for its size for numerical calculations. For calculations, which will be discussed in section \ref{sec:numeric}, sufficient accuracy was achieved for the matrix of size $10 \times 10$.

\section{The homogeneous solution and corrections to it}
Let us study the properties of the matrix $\hat K$. The matrix elements can be divided in to two contributions as follows:
\begin{equation}\label{eq:Kmn}
\begin{aligned}
K_{mn}(T_c) &= \int\limits_0^{\omega_D} \Re\left\{\frac{2}{i D h_n} \right\} \tanh\left(\frac{\varepsilon}{2 T_c} \right) d \varepsilon \int\limits_0^d \psi_n(x) \psi_m(x) dx +\\
&+\int\limits_0^{\omega_D} \tanh\left(\frac{\varepsilon}{2 T_c} \right) d \varepsilon \int\limits_0^d \Re \left\{\lambda_n^+(\varepsilon) e^{k x} + \lambda_n^-(\varepsilon) e^{-k x} \right\} \psi_m(x) dx
\end{aligned}
\end{equation} 
The first term is diagonal due to orthogonality of functions $\psi_n(x)$. Substituting $h_n(\varepsilon)$ from (\ref{eq:psi}) and taking the real part we find it to be
\begin{equation}
K^0_{mn}(T_c) = \delta_{mn} \int\limits_0^{\omega_D} \frac{4 \varepsilon d^4}{4 d^4 \varepsilon^2 + \pi^4 n^4 D^2} \tanh\left(\frac{\varepsilon}{2 T_c} \right) d \varepsilon.
\end{equation}
Here $\delta_{mn}$ is the Kronecker delta symbol. As we will see, it contributes to the homogeneous solution of the self--consistency equation. The second term leads to corrections due to influence of the borders of the superconductor.

\subsection{Homogeneous solution}
The homogeneous solution $\Delta(x) = \operatorname{const}$ should correspond to homogeneous boundary conditions $\alpha_0 = \alpha_d = 0$ in (\ref{eq:BC}), let us check that this is indeed the case. The homogeneous boundary conditions lead to $\lambda_n^\pm(\varepsilon) = 0$ in (\ref{eq:f_sol}) (see Appendix \ref{sec:appLambda}) and consequently the second term in (\ref{eq:Kmn}) is zero. So, the eigenproblem for $\hat K$ reduces to
\begin{equation}
c_m = \frac{U}{2} \sum_n K^0_{mn}(T_c) c_m
\end{equation}
and can be easily solved. The corresponding characteristic equation is
\begin{equation}
\left|\hat K^0(T_c) - \frac{2}{U} \right| = \prod_n \left(K^0_{nn}(T_c) - \frac{2}{U} \right) = 0.
\end{equation} 
The homogeneous solution, such that $\Delta(x) = \operatorname{const}$, corresponds to
\begin{equation}\label{eq:K00}
K^0_{00}(T_c) = \int\limits_0^{\omega_D} \frac{1}{2 \varepsilon} \tanh\left(\frac{\varepsilon}{2 T_c} \right) d \varepsilon = \frac{1}{U}
\end{equation}
and $c_n = \delta_{n0}$. In this case the only function in the expansion (\ref{eq:Delta_expansion}) is $\Phi_0(x)$ and $\Phi_0(x) \sim \psi_0(x) = \operatorname{const}$, given that $\lambda_n^\pm = 0$. The equation (\ref{eq:K00}) is a self--consistency equation for a bulk superconductor, which follows from the BCS theory \cite{tinkham_introduction_2015}.

\subsection{Corrections to homogeneous solution}
If the coefficients $\alpha_{0,d}(\varepsilon)$ are small (i.e. there is a small parameter in their definition), one can employ standard methods of perturbation theory in order to find corrections to the homogeneous solution, treating the matrix $\hat K^1$ with matrix elements
\begin{equation}
K^1_{mn}(T_c) = \int\limits_0^{\omega_D} \tanh\left(\frac{\varepsilon}{2 T_c} \right) d \varepsilon \int\limits_0^d \Re \left\{\lambda_n^+(\varepsilon) e^{k x} + \lambda_n^-(\varepsilon) e^{-k x} \right\} \psi_m(x) dx
\end{equation}
as a perturbation. For example, the first order correction shifts the critical temperature $T_c$ modifying the equation (\ref{eq:K00}) as
\begin{equation}
\begin{aligned}
&K^0_{00}(T_c) + K^1_{00}(T_c) = \int\limits_0^{\omega_D} \left[\frac{1}{\varepsilon} + g(\varepsilon) \right] \tanh\left(\frac{\varepsilon}{2 T_c} \right) d \varepsilon = \frac{2}{U}\\
&g(\varepsilon) = \int\limits_0^d \Re \left\{\lambda_0^+(\varepsilon) e^{k x} + \lambda_0^-(\varepsilon) e^{-k x} \right\} dx.
\end{aligned}
\end{equation}
The wave vector $k$ is defined by (\ref{eq:Usadel}). In the second order one finds corrections to coefficients $c_n$:
\begin{equation}
\begin{aligned}
c_n &\approx \delta_{n0} + \sum_{m \neq n} \frac{K^1_{mn}}{K^0_{nn} - K^0_{mm}} \delta_{m0} = \delta_{n0} + (1 - \delta_{n0}) \frac{K^1_{0n}}{K^0_{nn} - K^0_{00}}\\
& = \begin{cases}
1, &n = 0\\
\dfrac{K^1_{0n}}{K^0_{nn} - K^0_{00}}, &n \neq 0.
\end{cases}
\end{aligned}
\end{equation}
This defines spatial distribution of the order parameter in the superconductor, which will now be nonhomogeneous due to nonzero $\lambda_n^\pm(\varepsilon)$ in functions $\Phi_n(x)$ in the expansion (\ref{eq:Delta_expansion}) and contributions from functions with $n > 0$.

\section{A  superconductor with proximity effect at a single border}
We illustrate the method with the example of a junction, composed of a superconductor in contact with another material at $x = l$ and with free border at $x = 0$ (which means contact with a non magnetic insulator). In this case $\alpha_0 = 0$ and $\alpha_d \neq 0$. As it will be demonstrated further in this section, the analytical calculations in the first order of the perturbation theory lead to the results, obtained previously in \cite{fominov_nonmonotonic_2002}. This serves as the validation of the method. So, solving the Usadel equation (\ref{eq:Usadel}) we find
\begin{equation}
\begin{aligned}
f_s(x) &= \frac{2}{i D} \left[ \int\limits_0^{l} H(x, x'; \varepsilon) \Delta(x') dx' - \right.\\
&\left.- \frac{\alpha_d(\varepsilon) \cosh(k x)}{\alpha_d(\varepsilon) \cosh(k d) - k \sinh(k d)}\int\limits_0^{l} H(d, x'; \varepsilon) \Delta(x') dx' \right] = \\
&= \frac{2}{i D}\sum_n \frac{1}{h_n} \left[\psi_n(x) - \frac{\alpha_d(\varepsilon) \cosh(k x)}{\alpha_d(\varepsilon) \cosh(k d) - k \sinh(k d)} \psi_n(d)  \right] \int\limits_0^d \psi_n(x') \Delta(x') dx'.
\end{aligned}
\end{equation}
\subsection{Exact numerical solution}\label{sec:numeric}
{In order to find the critical temperature numerically, we write down the matrix elements of the matrix $\hat K(T_c)$, which are given by (\ref{eq:cK}). With all functions substituted, in the considered case they are given by
\begin{equation}
\begin{aligned}
K_{mn}(T_c) &= \int\limits_0^d dx \int\limits_0^{\omega_D} \psi_m(x) \Re \left\{\frac{2}{i D h_n} \left[\psi_n(x) - \frac{\alpha_d(\varepsilon) \cosh(k x)}{\alpha_d(\varepsilon) \cosh(k d) - k \sinh(k d)} \psi_n(d)  \right] \right\} \times\\
&\times \tanh \left(\frac{\varepsilon}{2T_c} \right) d \varepsilon.
\end{aligned}
\end{equation}
The integral over $x$ can be calculated analytically and we find for the matrix elements
\begin{equation}
\begin{aligned}
K_{mn}(T_c) &=  \int\limits_0^{\omega_D} \Re \left\{\frac{2}{i D h_n} \left[\delta_{nm} - \frac{2 \alpha_d(\varepsilon)}{\alpha_d(\varepsilon) \cosh(k d) - k \sinh(k d)} \frac{(-1)^{m+n} k d \sinh(k d)}{\pi^2 m^2 + k^2 d^2} \right] \right\} \times\\
&\times \tanh \left(\frac{\varepsilon}{2T_c} \right) d \varepsilon.
\end{aligned}
\end{equation}
Also the dependence $\alpha_d(\varepsilon)$ has to be specified. We choose
\begin{equation}\label{eq:alpha_eps}
\alpha_d(\varepsilon) = -\frac{c_1 \sqrt{2 i \varepsilon}}{1 + c_2 \sqrt{2 i \varepsilon}},
\end{equation}
it describes the proximity effect due to semiinfinite normal metal, see Appendix \ref{sec:appBC} for derivation. The dimension of $c_1$ is inverse length and $c_2$ is dimensionless, thus $c_1$ might be understood as the strength of the proximity effect and $c_2$ as the correction at different energy scales.

Next the critical temperature is found by performing numerical integration and solving the equation
\begin{equation}
\left| \hat K(T_c) - \frac{2}{U} \right| = 0.
\end{equation}
We plot the dependences of the critical temperature $T_c$ on the constant $c_1$ for different thickness $l$, the plots are presented in fig.\ref{fig:Tc_alpha}. As one can see, the critical temperature decreases as the proximity effect strength increases and turns to zero at some critical value of $c_1$. Also the reduction of $T_c$ is greater for thinner superconductor. The steep decline of the plots at critical values of the proximity effect strength $c_1$ is a sign of the first order phase transition \cite{fominov_nonmonotonic_2002, radovic_transition_1991}.
\begin{figure}[h!!]
\center\includegraphics[width = 0.5\textwidth]{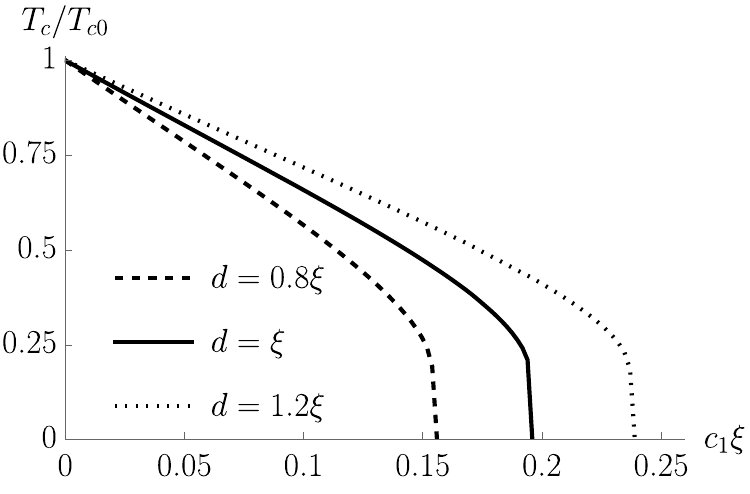}
\caption{Dependences of the critical temperature of the superconductor of thickness $d$ on the proximity effect strength $c_1$ obtained by numerical means. The thickness is normalized to the coherence length $\xi = \sqrt{D/(2 \pi T_{c0})}$, where $T_{c0}$ is the bulk superconductor critical temperature, and $c_2 = 1$.}
\label{fig:Tc_alpha}
\end{figure}

\subsection{Approximate homogeneous solution}
The homogeneous solution can be studied by finding the matrix element $K_{00}$ and subsequent self--consistency equation:
\begin{equation}
K_{00} = \int\limits_0^{\omega_D} \frac{1}{\varepsilon} \tanh\left(\frac{\varepsilon}{2 T_c} \right) d \varepsilon + \int\limits_0^{\omega_D} \Re \left\{\frac{4}{i D k^2} \frac{\alpha_d(\varepsilon)}{k^2 d - \alpha_d(\varepsilon) k d \coth(k d)} \right\} \tanh\left(\frac{\varepsilon}{2 T_c} \right) d \varepsilon = \frac{2}{U}.
\end{equation}
Here the first term corresponds to the self--consistency equation in the absence of the proximity effect and the second term --- the first order correction to it. We restrict ourself only with this first order perturbation, which means that there are no spatial corrections to the order parameter. This approximation is valid for the thin superconductor in which $\Delta(x)$ is constant. Qualitatively, this means $d \ll \xi = \sqrt{D/(2\pi T_{c0})}$ and in this case
\begin{equation}
\frac{\alpha_d(\varepsilon)}{k^2 d - \alpha_d(\varepsilon) k d \coth(k d)}  \approx \frac{\alpha_d(\varepsilon)}{k^2 d - \alpha_d(\varepsilon)} .
\end{equation}
Also we consider the constant $\alpha_d(\varepsilon)$, which is achieved for large $c_1$ and $c_2$:
\begin{equation}\label{eq:alpha_d}
\alpha_d(\varepsilon) \approx -\frac{c_1}{c_2} = \operatorname{const}.
\end{equation}
From physical point of view, this corresponds to small diffusion constant $D_n$ in the adjacent normal metal or large conductivity of the normal metal, see equation (\ref{eq:c12}). Then
\begin{equation}
\Re \left\{\frac{4}{i D k^2} \frac{\alpha_d}{k^2 d - \alpha_d k d \coth(k d)} \right\} \approx \Re \left\{\frac{4}{i D k^2} \frac{\alpha_d}{k^2 d - \alpha_d} \right\} = \frac{2 \alpha_d^2 D^2}{ \alpha_d^2 D^2 \varepsilon + 4 d^2 \varepsilon^3}.
\end{equation}
Finally, the self--consistency equation becomes
\begin{equation}
\int\limits_0^{\omega_D} \left\{\frac{1}{2 \varepsilon} + \frac{\alpha_d^2 D^2}{ \alpha_d^2 D^2 \varepsilon + 4 d^2 \varepsilon^3} \right\} \tanh\left(\frac{\varepsilon}{2 T_c} \right) d \varepsilon = \frac{1}{U}.
\end{equation}
The integral, given by the second term, converges in the infinite limits and we can remove the cutoff $\omega_D$ in the first term using the standard regularization procedure. This gives
\begin{equation}
\ln \left(\frac{T_{c0}}{T_c} \right) = \int\limits_0^\infty  \frac{1}{\varepsilon + \sigma^2 \varepsilon^3/T_{c0}^2}\tanh\left(\frac{\varepsilon}{2 T_c} \right) d \varepsilon = \int\limits_0^\infty  \frac{1}{s + \sigma^2 s^3}\tanh\left(\frac{s}{2 } \frac{T_{c0}}{T_c} \right) d s,
\end{equation}
where $T_{c0}$ is the critical temperature of the bulk superconductor and we have introduced the dimensionless parameter
\begin{equation}
\sigma = \frac{d}{\pi |\alpha_d| \xi^2}.
\end{equation}
The integral can be evaluated exactly using Matsubara summation. First we expand the hyperbolic tangent as
\begin{equation}
\begin{aligned}
&\tanh\left(\frac{s}{2 } \frac{T_{c0}}{T_c} \right) = \frac{4 s T_c}{T_{c0}}\sum_{\omega_n} \frac{1}{s^2 + \omega_n^2/T_{c0}^2}\\
&\omega_n = (2n + 1)\pi T_c,\ n > 0.
\end{aligned}
\end{equation}
Then the integral can be represented as a sum:
\begin{equation}
\begin{aligned}
\int\limits_0^\infty  \frac{1}{s + \sigma^2 s^3}\tanh\left(\frac{s}{2 } \frac{T_{c0}}{T_c} \right) d s &= \frac{T_c}{T_{c0}}\sum_{\omega_n} \int\limits_0^\infty \frac{4s}{(s + \sigma^2 s^3)(s^2 + \omega_n^2/T_{c0}^2)} ds =\\
&= \sum_{\omega_n} \frac{2\pi T_c}{\omega_n + \sigma \omega_n^2/T_{c0}}.
\end{aligned}
\end{equation}
The evaluation of the sum leads to the self--consistency equation
\begin{equation}
\ln \left(\frac{T_{c0}}{T_c} \right) = \psi\left(\frac{1}{2} + \frac{T_{c0}}{2 \pi \sigma T_c} \right) - \psi \left(\frac{1}{2} \right),
\end{equation}
where $\psi(x)$ is the digamma function. With all dimensional physical quantities substituted, see equations (\ref{eq:alpha_d}) and (\ref{eq:c12}), this is exactly the self--consistency equation (B5) in ref. \cite{fominov_nonmonotonic_2002}, which was obtained in the single--mode approximation. The solution of this equation is the dependence of the critical temperature $T_c(\sigma)$, which is numerically plotted in fig. \ref{fig:Tc}.
\begin{figure}[h!!]
\center\includegraphics[width = 0.5\textwidth]{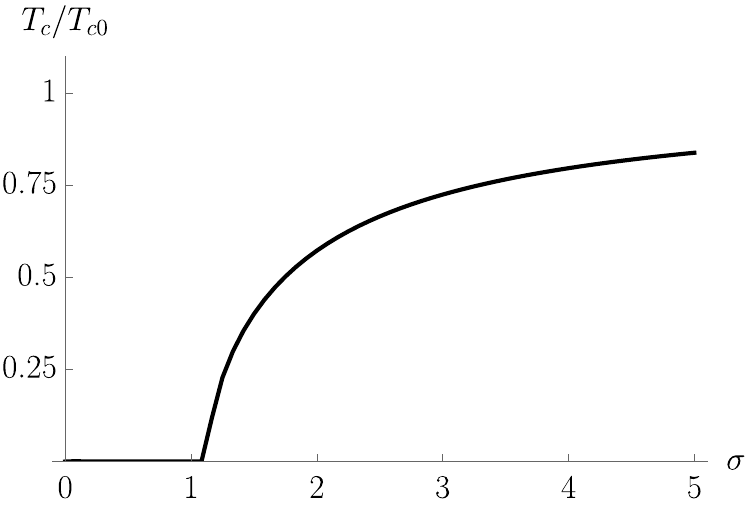}
\caption{Dependence of the critical temperature of the thin superconductor with the proximity effect at a single border on the parameter $\sigma = d/(\pi |\alpha_d| \xi^2)$. The critical temperature turns to zero at $\sigma_c = 2 e^\gamma/\pi \approx 1.12$.}
\label{fig:Tc}
\end{figure}

From the condition $T_c = 0$ we find the critical value $\sigma_c = 2 e^\gamma/\pi \approx 1.12$, where $\gamma \approx 0.58$ is the Euler constant. This gives the critical thickness of the junction $(d_c/\xi) = \pi \sigma_c |\alpha_d| \xi \approx 3.56 |\alpha_d| \xi$. Also the critical temperature $T_c$ of the layer tends to the critical temperature $T_{c0}$ of the bulk superconductor in the limit of $\sigma \rightarrow \infty$. This is the limit of the weak proximity effect (or one might consider the junction thick, although technically we assumed it was thin enough in our derivation).

In \cite{bakurskiy_theoretical_2013} the critical thickness was calculated from the linearized Usadel equation in another limit $\xi \ll d \ll \xi(T)$, where the last value $\xi(T)$ is the Ginzburg--Landau coherence length, while $\xi$ is the microscopic coherence length: $\xi(T)= \ (\pi\xi/2) \sqrt{1-T/T_c}$, and assuming a strong suppression of superconductivity at the border: $\Delta(d)=0$. Then the critical thickness was found as $d_c = \pi \xi(T)/2$.

\section{Discussion}
We have carried out analytical study of the coupled linearised Usadel equation and the self--consistency equation. The approach is based on expressing the solution of the Usadel equation in integral form and then substituting it in the self--consistency equation, which gives an integral equation for the order parameter \cite{fominov_nonmonotonic_2002, antropov_experimental_2013}. The integral equation was reduced to an eigenvalue problem for a certain matrix by expanding the integral kernel in the basis of the eigenfunctions of the differential operator, corresponding to the Usadel equation. The eigenvalue problem might be solved numerically in the general case and also further analytical study might be possible. Its solution provides the value for the critical temperature of the superconductor and the spatial dependence of the order parameter. In particular we have considered the case of a superconductor in the presence of the proximity effect with a semiinfinite normal metal at a single border. In the first order perturbation theory, which is sufficient for the short superconductor, we have reproduced the single--mode approximation self--consistency equation from \cite{fominov_nonmonotonic_2002} and numerically solved it. The solution provides dependence of the critical temperature on the strength of the proximity effect and thickness of the superconductor. Also the exact diagonalization was performed numerically for the general case, dependences of the critical temperature on the strength of the proximity effect for different thickness of the junction were obtained.  

\section{Acknowledgments}
The publication was prepared within the framework of the Academic Fund Program at HSE University \#24-00-038 ``Quasiclassical dynamics of quantum systems: chaotic and spatially nonhomogeneous like heterostructures superconductor--magnetic''.


\begin{thebibliography}{10}

\bibitem{de_gennes_boundary_1964}
P.~G. De~Gennes, ``Boundary {Effects} in {Superconductors},'' {\em Reviews of
  Modern Physics}, vol.~36, pp.~225--237, Jan. 1964.

\bibitem{buzdin_proximity_2005}
A.~I. Buzdin, ``Proximity effects in superconductor-ferromagnet
  heterostructures,'' {\em Reviews of Modern Physics}, vol.~77, pp.~935--976,
  Sept. 2005.

\bibitem{eschrig_spin-polarized_2015}
M.~Eschrig, ``Spin-polarized supercurrents for spintronics: a review of current
  progress,'' {\em Reports on Progress in Physics}, vol.~78, p.~104501, Oct.
  2015.

\bibitem{bergeret_odd_2005}
F.~S. Bergeret, A.~F. Volkov, and K.~B. Efetov, ``Odd triplet superconductivity
  and related phenomena in superconductor-ferromagnet structures,'' {\em
  Reviews of Modern Physics}, vol.~77, pp.~1321--1373, Nov. 2005.

\bibitem{eschrig_general_2015}
M.~Eschrig, A.~Cottet, W.~Belzig, and J.~Linder, ``General boundary conditions
  for quasiclassical theory of superconductivity in the diffusive limit:
  application to strongly spin-polarized systems,'' {\em New Journal of
  Physics}, vol.~17, p.~083037, Aug. 2015.

\bibitem{Leijnse_2012}
M.~Leijnse and K.~Flensberg, ``Introduction to topological superconductivity
  and majorana fermions,'' {\em Semiconductor Science and Technology}, vol.~27,
  p.~124003, nov 2012.

\bibitem{hassler_majorana_2014}
F.~Hassler, ``Majorana {Qubits},'' 2014.

\bibitem{seleznev_density_2025}
D.~V. Seleznev, S.~S. Seidov, N.~G. Pugach, D.~G. Bezymiannykh, S.~I. Mukhin,
  and B.~G. L’vov, ``Density of {States} in the {Heterostructure}
  {Ferromagnetic} {Insulator}-{Superconductor}-{Ferromagnetic} {Insulator},''
  {\em Journal of Superconductivity and Novel Magnetism}, vol.~38, p.~9, Feb.
  2025.

\bibitem{linder_superconducting_2015}
J.~Linder and J.~W.~A. Robinson, ``Superconducting spintronics,'' {\em Nature
  Physics}, vol.~11, pp.~307--315, Apr. 2015.

\bibitem{leksin_peculiarities_2013}
P.~V. Leksin, A.~A. Kamashev, N.~N. Garif’yanov, I.~A. Garifullin, Y.~V.
  Fominov, J.~Schumann, C.~Hess, V.~Kataev, and B.~Büchner, ``Peculiarities of
  performance of the spin valve for the superconducting current,'' {\em JETP
  Letters}, vol.~97, pp.~478--482, June 2013.

\bibitem{leksin_boosting_2016}
P.~V. Leksin, A.~A. Kamashev, J.~Schumann, V.~E. Kataev, J.~Thomas,
  B.~Büchner, and I.~A. Garifullin, ``Boosting the superconducting spin valve
  effect in a metallic superconductor/ferromagnet heterostructure,'' {\em Nano
  Research}, vol.~9, pp.~1005--1011, Apr. 2016.

\bibitem{kamashev_superconducting_2019}
A.~A. Kamashev, N.~N. Garif'yanov, A.~A. Validov, J.~Schumann, V.~Kataev,
  B.~Büchner, Y.~V. Fominov, and I.~A. Garifullin, ``Superconducting
  spin-valve effect in heterostructures with ferromagnetic {Heusler} alloy
  layers,'' {\em Physical Review B}, vol.~100, p.~134511, Oct. 2019.

\bibitem{gordeeva_record_2020}
A.~V. Gordeeva, A.~L. Pankratov, N.~G. Pugach, A.~S. Vasenko, V.~O. Zbrozhek,
  A.~V. Blagodatkin, D.~A. Pimanov, and L.~S. Kuzmin, ``Record electron
  self-cooling in cold-electron bolometers with a hybrid
  superconductor-ferromagnetic nanoabsorber and traps,'' {\em Scientific
  Reports}, vol.~10, p.~21961, Dec. 2020.

\bibitem{heim_ferromagnetic_2013}
D.~M. Heim, N.~G. Pugach, M.~Y. Kupriyanov, E.~Goldobin, D.~Koelle, and
  R.~Kleiner, ``Ferromagnetic planar {Josephson} junction with transparent
  interfaces: a $\varphi$ junction proposal,'' {\em Journal of Physics:
  Condensed Matter}, vol.~25, p.~215701, May 2013.

\bibitem{usadel_generalized_1970}
K.~D. Usadel, ``Generalized {Diffusion} {Equation} for {Superconducting}
  {Alloys},'' {\em Physical Review Letters}, vol.~25, pp.~507--509, Aug. 1970.

\bibitem{tinkham_introduction_2015}
M.~Tinkham, {\em Introduction to {Superconductivity}: {Second} {Edition}}.
\newblock Mineola, NY: Dover Publications, 2015.

\bibitem{tumanov_2024}
V.~Tumanov and Y.~Proshin, ``Comparison of several methods for determining the
  critical temperature of a superconducting transition in
  ferromagnet/superconductor heterostructures,'' {\em Magnetic resonance in
  solids}, vol.~26, no.~1, 2024.

\bibitem{ivanov_boundary_1981}
Z.~G. Ivanov, M.~Y. Kupriyanov, K.~K. Likharev, S.~V. Meriakri, and O.~V.
  Snigirev, ``Boundary conditions for the {Eilenberger} and {Usadel} equations
  and properties of “dirty” {SNS} sandwiches,'' {\em Soviet Journal of Low
  Temperature Physics}, vol.~7, pp.~274--281, May 1981.

\bibitem{fominov_nonmonotonic_2002}
Y.~V. Fominov, N.~M. Chtchelkatchev, and A.~A. Golubov, ``Nonmonotonic critical
  temperature in superconductor/ferromagnet bilayers,'' {\em Physical Review
  B}, vol.~66, p.~014507, June 2002.

\bibitem{antropov_experimental_2013}
E.~Antropov, M.~S. Kalenkov, J.~Kehrle, V.~I. Zdravkov, R.~Morari,
  A.~Socrovisciuc, D.~Lenk, S.~Horn, L.~R. Tagirov, A.~D. Zaikin, A.~S.
  Sidorenko, H.~Hahn, and R.~Tidecks, ``Experimental and theoretical analysis
  of the upper critical field in ferromagnet–superconductor–ferromagnet
  trilayers,'' {\em Superconductor Science and Technology}, vol.~26, p.~085003,
  Aug. 2013.

\bibitem{zaitsev_quasiclassical_1984}
A.~Zaitsev, ``Quasiclassical equations of the theory of superconductivity for
  contiguous metals and the properties of constricted microcontacts,'' {\em
  Soviet Physics - JETP}, vol.~59, no.~5, pp.~1015--1024, 1984.
\newblock INIS Reference Number: 16069709.

\bibitem{kurpianov_influence_1988}
M.~Kuprianov and V.~Lukichev, ``Influence of boundary transparency on the
  critical current of dirty {SS}'{S} structures,'' {\em Soviet Physics - JETP
  (English Translation)}, vol.~67, no.~6, pp.~1163--1168, 1988.
\newblock INIS Reference Number: 22024754.

\bibitem{linder_quasiclassical_2022}
J.~Linder and M.~Amundsen, ``Quasiclassical boundary conditions for spin-orbit
  coupled interfaces with spin-charge conversion,'' {\em Physical Review B},
  vol.~105, p.~064506, Feb. 2022.

\bibitem{bergeret_spin-orbit_2014}
F.~S. Bergeret and I.~V. Tokatly, ``Spin-orbit coupling as a source of
  long-range triplet proximity effect in superconductor-ferromagnet hybrid
  structures,'' {\em Physical Review B}, vol.~89, p.~134517, Apr. 2014.

\bibitem{kokkeler_universal_2025}
T.~Kokkeler, F.~S. Bergeret, and I.~Tokatly, ``A universal phenomenology of
  charge-spin interconversion and dynamics in diffusive systems with spin-orbit
  coupling,'' Jan. 2025.
\newblock arXiv:2405.06334.

\bibitem{duffy_greens_2001}
D.~G. Duffy, {\em Green's {Functions} with {Applications}}.
\newblock Boca Raton, Fla. London: Chapman and Hall/CRC, 2001.

\bibitem{polyanin_handbook_1998}
A.~D. Polyanin and A.~V. Manzhirov, {\em Handbook of {Integral} {Equations}}.
\newblock Boca Raton, Fla.: CRC Press, 1998.

\bibitem{radovic_transition_1991}
Z.~Radović, M.~Ledvij, L.~Dobrosavljević-Grujić, A.~I. Buzdin, and J.~R.
  Clem, ``Transition temperatures of superconductor-ferromagnet
  superlattices,'' {\em Physical Review B}, vol.~44, pp.~759--764, July 1991.

\bibitem{bakurskiy_theoretical_2013}
S.~V. Bakurskiy, N.~V. Klenov, I.~I. Soloviev, V.~V. Bol'ginov, V.~V. Ryazanov,
  I.~V. Vernik, O.~A. Mukhanov, M.~Y. Kupriyanov, and A.~A. Golubov,
  ``Theoretical model of superconducting spintronic {SIsFS} devices,'' {\em
  Applied Physics Letters}, vol.~102, p.~192603, May 2013.

\end{thebibliography}

\appendix
\section{Functions $\lambda^\pm_n(\varepsilon)$}\label{sec:appLambda}
In this section we derive functions $\lambda_n^\pm(\varepsilon)$ in (\ref{eq:apm}), which are related to the influence of the boundary. First we substitute the solution (\ref{eq:f_sol}) of the Usadel equation in the boundary conditions (\ref{eq:BC}) and find
\begin{equation}
a_\pm(\varepsilon) = \mu_\pm(\varepsilon) \int\limits_0^d H(0, x'; \varepsilon) \Delta(x') dx' + \nu_\pm(\varepsilon) \int\limits_0^d H(d, x'; \varepsilon) \Delta(x') dx'.
\end{equation}
The coefficients $\mu_\pm(\varepsilon)$ and $\nu_\pm(\varepsilon)$ are
\begin{equation}
\begin{aligned}
&\begin{aligned}
&\mu_+(\varepsilon) = \frac{\alpha_0}{i D s} (\alpha_d + k)& &\nu_+(\varepsilon) = \frac{\alpha_d e^{k d}}{i D s} (\alpha_0 + k)\\
&\mu_-(\varepsilon) = \frac{\alpha_0 e^{2 k d}}{i D s} (\alpha_d - k)& &\nu_-(\varepsilon) = \frac{\alpha_d e^{2 k d}}{i D s}  (k - \alpha_0)
\end{aligned}\\
&s = e^{2 k d} \big(\alpha_0 + k \big) \big(\alpha_d - k \big) + \big(\alpha_0 - k \big) \big(\alpha_d + k \big).
\end{aligned}
\end{equation}
Next we use expansion (\ref{eq:H0_eigen}) of the Green function $H(x, x'; \varepsilon)$ from which follows
\begin{equation}
a_\pm(\varepsilon) = \sum_n \frac{\mu_\pm(\varepsilon) + (-1)^n \nu_\pm (\varepsilon)}{h_n}  \int\limits_0^d \psi_n(x') \Delta(x') dx'.
\end{equation}
Comparing with (\ref{eq:apm}) we find
\begin{equation}\label{eq:lambda}
\lambda_n^\pm(\varepsilon) = \frac{\mu_\pm(\varepsilon) + (-1)^n \nu_\pm (\varepsilon)}{h_n}.
\end{equation}

\section{Effective boundary conditions for the superconductor in contact with a semiinfinite normal metal}\label{sec:appBC}
The boundary conditions at the superconductor--normal metal interface are the Kuprianov--Lukichev conditions \cite{kurpianov_influence_1988}
\begin{equation}\label{eq:BC_KupLuk}
\begin{aligned}
&f'_s(d) = \gamma f'_N(d)\\
&f'_n(d) = \beta \left[f_s(d) - f_n(d) \right].
\end{aligned}
\end{equation}
The anomalous Green function in the normal metal $f_n(x)$ obeys the Usadel equation
\begin{equation}
\begin{aligned}
&f''_n(x) - k_n^2 f_n(x) = 0\\
&k_n^2 = \frac{2 i \varepsilon}{D_n}.
\end{aligned}
\end{equation}
Here $D_n$ is the diffusion constant in the normal metal. The coefficients $\gamma$ and $\beta$ are expressed via the conductivities of the superconductor and the normal metal $\sigma_s$ and $\sigma_n$ respectively, and the specific interface resistance $R$ as
\begin{equation}
\begin{aligned}
&\gamma = \frac{\sigma_n}{\sigma_s}\\
&\beta = \frac{1}{\sigma_n R}.
\end{aligned}
\end{equation} 

We consider the case of the semiinfinite normal metal, so $f_n(\infty) = 0$ and the solution of the normal metal Usadel equation is
\begin{equation}
f_n(x) = C e^{-k_n x}.
\end{equation}
Thus we find $f'_n(d) = -k_n f_n(d)$ and the boundary conditions (\ref{eq:BC_KupLuk}) become
\begin{equation}
\begin{aligned}
&f'_s(d) = -\gamma k_n f_n(d)\\
-&k_n f_n(d) = \beta \left[f_n(d) - f_s(d) \right].
\end{aligned}
\end{equation}
This can be solved for $f'_s(d)$ and we find
\begin{equation}
f'_s(d) = -\frac{\beta \gamma k_n}{\beta + k_n} f_s(d) = -\frac{\gamma}{\sqrt{D_n}}\frac{\sqrt{2 i \varepsilon}}{1 + \sqrt{2 i \varepsilon}/(\beta \sqrt{D_n})} f_s(d).
\end{equation}
This relates to the coefficient $\alpha_d(\varepsilon)$ in equation (\ref{eq:alpha_eps}) if one sets
\begin{equation}\label{eq:c12}
\begin{aligned}
&c_1 = \frac{\gamma}{\sqrt{D_n}} = \frac{\sigma_n}{\sigma_s \sqrt{D_n}}\\
&c_2 = \frac{1}{\beta \sqrt{D_n}} = \frac{\sigma_n R}{\sqrt{D_n}}.
\end{aligned}
\end{equation}
\end{document}